\documentstyle[12pt]{article}

\newcommand{\nc}{\newcommand}
\nc{\beq}{\begin{equation}}
\nc{\eeq}{\end{equation}}
\nc{\beqa}{\begin{eqnarray}}
\nc{\eeqa}{\end{eqnarray}}

\input{epsf.sty}  
\input{psfig.sty}
\newwrite\ffile\global\newcount\figno \global\figno=1

\def\writedef#1{}
\def\figin{\epsfcheck\figin}\def\figins{\epsfcheck\figins}
\def\epsfcheck{\ifx\epsfbox\UnDeFiNeD
\message{(NO epsf.tex, FIGURES WILL BE IGNORED)}
\gdef\figin##1{\vskip2in}\gdef\figins##1{\hskip.5in}
\else\message{(FIGURES WILL BE INCLUDED)}%
\gdef\figin##1{##1}\gdef\figins##1{##1}\fi}
\def\figinsert{}
\def\ifig#1#2#3{\xdef#1{fig.~\the\figno}
\writedef{#1\leftbracket fig.\noexpand~\the\figno}%
\figinsert\figin{\centerline{#3}}\medskip\centerline{\vbox{\baselineskip12pt
\advance\hsize by -1truein\center\footnotesize{  Fig.~\the\figno.} #2}}
\bigskip\endinsert\global\advance\figno by1}
\def\endinsert{}

\begin{document}

\title{\large{\bf Magnetic Interactions, the Renormalization Group 
and Color Superconductivity in High Density QCD}}

\author{ 
Stephen D.H.~Hsu\thanks{hsu@duende.uoregon.edu}
 \\ {\small Department of Physics, 
University of Oregon, Eugene OR 97403-5203.} \\ \\
Myckola Schwetz\thanks{myckola@baobab.rutgers.edu}
 \\ {\small Department of Physics and Astronomy, 
Rutgers University, 
Piscataway NJ 08855-0849.}\\ \\
}

\date{August 1999}

\maketitle

\begin{picture}(0,0)(0,0)
\put(350,360){OITS-676}
\put(350,340){RU-99-25}
\end{picture}
\vspace{-24pt}

\begin{abstract}
We investigate the effect of long range magnetic interactions on the 
renormalization group (RG) evolution of local Cooper pairing interactions
near the Fermi surface in high density QCD. We use an explicit 
cut-off on momentum modes, with special emphasis on screening effects 
such as Landau damping, to derive the RG equations in a gauge invariant,
weak coupling expansion. We obtain the Landau pole  
$\Delta \sim \mu g^{-5} \exp( - \frac{ 3 \pi^2}{ \sqrt{2} g})~$, although
the structure of our equations differs from previous results.
We also investigate the gap equation, including condensates of higher angular 
momentum. We show that rotational invariance is unbroken at asymptotically 
high density, and verify that $\Delta$ is the correct value of the gap when
higher modes are included in the analysis.

\end{abstract}

\newpage

\section{Introduction}

In this paper we study the behavior of quark matter at high density
and low temperature \cite{colorsuper}-\cite{PRTc}. Under these conditions
QCD exhibits color superconductivity, caused by the condensation of
diquarks. This condensation is analogous to the Cooper pairing observed
in ordinary superfluids, and can be shown to occur in the presence of
even very weakly attractive interactions. The reason for this is the
special nature of physics very close to the Fermi surface (FS).

At high density, characterized by a chemical
potential $\mu$ which is much larger than the current quark masses and
the QCD scale $\Lambda_{QCD}$, the typical momentum transfer in 
quark--quark interactions is of order $\mu$, and therefore it is plausible
that the dynamics can be understood via perturbative gluon exchange.
Of course, small angle scatterings, which involve small momentum transfers,
are still problematic and require special attention.

In recent work, Son \cite{Son} showed that long range magnetic interactions
lead to a modification of the RG equations originally derived in \cite{EHS}
for the case of local interactions. The magnetic effects are strong enough
to modify the parametric dependence of the position of the Landau pole 
(and hence the superfluid gap $\Delta$) on the gauge coupling
constant. Son finds
\beq
\label{LP}
\Delta \sim \mu g^{-5} \exp \Bigl( - {3 \pi^2 \over \sqrt{2} g} \Bigr) ~~~.
\eeq
That $\Delta$ should scale like $\exp (-c/g)$ is easy to see \cite{PR}
by considering the gap equation with a massless, or weakly damped,
gauge propagator. In the usual case of a local four fermion interaction,
the gap integral exhibits a logarithmic divergence 
which, roughly speaking, is cut off
near the FS by the gap itself. Solving for $\Delta$ yields a result of the
form $\Delta \sim \exp(-c/G)$, where G is the four fermion coupling, and
is of order $g^2$ if it arises from the exchange of a gauge boson. However,
if the four fermion interaction is replaced by a weakly damped gauge
propagator, an additional logarithmic divergence appears due to small angle 
scattering of the fermions. This divergence is again regulated by a scale
related to the gap itself, and the resulting exponent of the solution is
roughly the square root of the what appeared in the local case:
$\Delta \sim \exp(-c/g)$. 

Our intention here is to understand this behavior in terms of the evolution
of Cooper pairing interactions near the Fermi surface. Despite the 
long range of the magnetic interactions, we find that the problem can still
be formulated in terms of local operators which are, essentially, terms in
the  expansion of the magnetic gluon form factor. The reason that this is
possible is because we retain at all times an explicit cutoff on long
wavelength modes, which keeps all quantities finite. This approach is somewhat
different from that of Son \cite{Son}, who studied the RG evolution of
scattering amplitudes themselves. As discussed below, our results differ
from his. Primarily, we believe that this is due to conceptual problems 
in applying the RG directly to scattering amplitudes.

The paper is organized as follows. In section 2 we give a description of
our cutoff scheme and the resulting effective lagrangian. In section 3 we
discuss the problem of small angle scattering and gluon screening effects.
In section 4 we compute our RG equations and compare them with Son's.
In section 5 we investigate the possibility of breaking rotational symmetry.
The final section contains additional discussion of our results.
 
\section{RG Scheme}

We adopt a Wilsonian RG procedure, with a hard IR cutoff on spatial momentum, 
$\Lambda$ \cite{EHS,Shankar}.
As $\Lambda \rightarrow 0$, only quark excitations very near the FS, as well
as soft gluons, are left in the effective theory.
Our prescription differs from what is often used in QFT, where a hard cut-off 
is imposed on energy as well as momentum, but it has some advantages. In 
particular, in our scheme integration over modes corresponds to shrinking
the Hilbert space of the model in the basis of energy eigenstates.

The effective Lagrangian has the form
\beq
{\cal L}_{\Lambda} ~=~ {\cal L}_{QCD} ~+~ \sum_n {\cal O}_n
\eeq
where the ${\cal O}_n$ are local operators involving quark and gluon
fields, which are the result of the integration over higher
frequency quarks and gluons. The most important of these operators 
are the marginal Cooper pairing interactions which involve
four quarks. All other quark interactions can be shown to be irrelevant in
the limit of small $\Lambda$ \cite{EHS}.

The Cooper pairing interaction is of the form
\beq
\label{G}
G(k-q) \bar{\psi}_+ (k_0, k) \gamma_{\mu} P_{L,R} \psi_+ (q_0,q) 
 \bar{\psi}_+ (k_0, -k) \gamma^{\mu} P_{L,R} \psi_+ (q_0, -q)~~~,
\eeq
where $\psi_+$ denotes the projection of the quark field 
\beq
\psi_+ = \frac{1}{2}  (1+ \vec{\alpha} \cdot \hat{p}) \psi (p)~~~,
\eeq
and consists of quark, rather than antiquark degrees of freedom.
We will be interested in the case where all of the external
quarks in this operator are essentially on-shell. 
Note that the incoming and outgoing quarks have almost equal and opposite
momenta. Near the FS, the form factor $G(k-q)$ becomes a function of
angle $\theta = {\vec{k} \cdot \vec{q} \over k q}$, since 
$|k| \simeq |q| \simeq \mu$, and $k_0, q_0 \simeq 0$. 
(Strictly speaking, in the case of Landau damping, 
it is a function of energy as well as angle. We will always
assume that the energy transfer $k_0 - q_0$ in the gluon line is 
much less than but of order $\Lambda$.)
In this paper we restrict ourselves to the $\bar{3}$ color channel, 
which is attractive and to the LL (or equivalently RR) chirality channel, 
which has been shown to dominate the LR channel \cite{EHS}. 
It is straightforward to derive the related 
RG equations for other channels using our techniques.

In previous work we took the form factor $G ( \theta )$ to be a constant
\cite{EHS}.
This is appropriate at sufficiently low energy, 
if screening masses exist for both the
magnetic and electric components of the gluon, which is likely to be
the case at intermediate densities where the coupling is not small. 
However, as argued by Son \cite{Son}, a magnetic mass for the gluon 
is unlikely to arise within
perturbation theory. The magnetic mass is due solely to nonperturbative
effects, and is presumably of order $\exp(-1/g^2)$. 
At high density it is therefore likely to be 
smaller than the eventual superfluid gap, and hence plays no role
in the analysis. Instead, \cite{Son} focused on the role of Landau
damping on the magnetic interactions. 

\begin{figure}[htb]
\center{
\leavevmode
\epsfbox{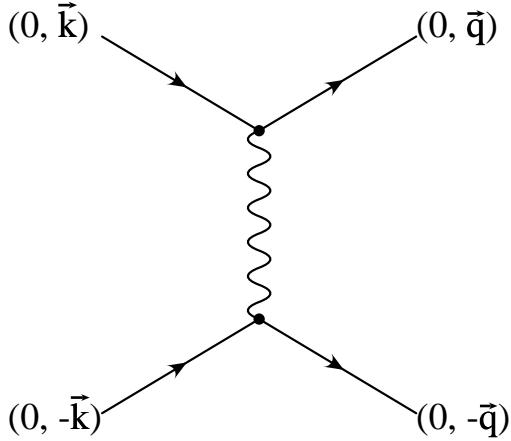}
\caption{One gluon exchange.} \label{gluon}
}
\end{figure}

In order to consider long range magnetic interactions, it is necessary
to expand the form factor $G ( \theta )$ in components with definite
angular momentum. We can then study the RG evolution of each of these
components. Let
\beq
G ( \theta ) ~=~ \sum_l (2l+1) P_l ( \cos \theta ) G_l ~~~,
\eeq
so that each component 
\beq
G_l = \frac{1}{2} \int_{-1}^{1} d (\cos \theta) ~ P_l (\cos \theta)
G ( \theta )~~~.
\eeq
This integral exhibits a logarithmic divergence 
in the case of a massless, or weakly damped, gluon (figure \ref{gluon}). 
However, in our regularization scheme $G(\theta)$
contains only the effects of gluons which have been integrated out
above the cutoff $\Lambda$. For nonzero cutoff the components
$G_l$ will be finite, but exhibit a logarithmic dependence on
$\Lambda$\footnote{The precise form of this logarithm is dependent
on the IR behavior of the propagator, as we will see below.}. 
It is this logarithmic dependence that leads to the
constant ${\cal O}(g^2)$ term in the RG equations noted by 
Son \cite{Son}.

\epsfysize=6 cm

\section{Screening Effects}

In this section we discuss the incorporation of screening effects
in our procedure. As mentioned previously, small angle scattering
of quarks must be considered carefully, as the simple perturbative
expansion in powers of $g^2$ may break down. Indeed, it is easy to
see that vacuum polarization corrections to any graph can become
important if the gluon momentum is sufficiently small. Resummation
of these effects leads to a screened propagator which is well behaved
at small momentum, as long as the energy is nonzero.

The gluon propagator, including vacuum
polarization effects from virtual quarks, \cite{Bellac} has the 
following form in covariant gauge:
\beq
\label{D}
D_{\mu\nu}~=~{1\over G + Q^2} P^T_{\mu\nu} ~+~ {1\over F + Q^2} P^L_{\mu\nu}
~-~ \xi  { Q_\mu Q_\nu \over Q^4}
~,
\eeq
where $Q = (q_4, \vec{q}) = (-\omega,\vec{q}) $ 
is the gluon Euclidean 4-momentum, and  $P^T_{\mu\nu}$
and $P^L_{\mu\nu}$ are transverse and longitudinal projectors correspondingly.
In our leading order calculations the propagator will always appear contracted
with gamma matrices next to on-shell external quark lines. Thus the
gauge dependent part of (\ref{D}) will vanish due to the equations of 
motion, leading to a gauge-invariant result. Henceforth we will simply set
$\xi = 0$ in (\ref{D}).

The functions $G$ and $F$ are related to the gluon polarization tensor
$\Pi_{\mu\nu}$:
\beqa
F~=~ {Q^2 \over q^2}\, \Pi_{44}~, 
~~~~~~~~~~~~~~~~~~~~~~~~~~~~~~~~ \nonumber \\
G~=~ {1\over2}\, P^T_{\mu\nu}~ \Pi_{\mu\nu}~=~ 
{1\over2}\,(\delta_{ij}~-~{q_i q_j \over q^2})\, \Pi_{ij}, \nonumber
\eeqa
which reflects the fact that the Lorentz symmetry is broken to 3D 
rotational symmetry.
The explicit expression of $\Pi_{\mu\nu}$ to one loop is
\beq
\label{Pi}
\Pi_{\mu\nu} (Q)~=~ g^2 \int {d^4K \over (2 \pi)^4} \, 
Tr [ \gamma_\mu \,{K\!\!\!\!/} \,\gamma_\nu \,({K\!\!\!\!/} - {Q\!\!\!\!/})
 ] \,\Delta(K) \Delta(K - Q)~,
\eeq
where $\Delta(K) = 1/K^2$. 

\epsfysize=6 cm
\begin{figure}[htb]
\center{
\leavevmode
\epsfbox{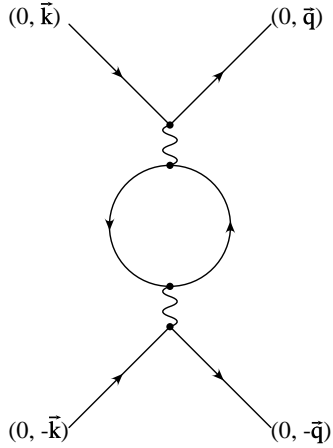}
\caption{Vacuum polarization correction to gluon propagator.} 
\label{polarization}
}
\end{figure}

If the energy and momentum transfer in figure (\ref{polarization})
are small then one may neglect
$Q$ in the numerator of (\ref{Pi}), as the dominant momenta in the loop will 
be $k \sim \mu$ (this is the equivalent of the 
hard thermal loop approximation).
Then $F$ and $G$ take the familiar form \cite{Bellac}
\beqa
\label{FG}
F~&=&~{2 m^2_D Q^2 \over q^2 }\, \Big( 1~-~{i \omega \over q}\,
    L_0 \Big({i \omega \over q }\Big) \Big)~ \nonumber \\
G~&=&~ m^2_D {i \omega \over q } 
\Big[ \Big( 1~-~\Big({i \omega \over q }\Big)^2
\Big) \, L_0 \Big({i \omega \over q }\Big) ~+~ {i \omega \over q } \Big]~ ,
\eeqa
where
\beq
L_0 (x)~=~ {1 \over 2}\, \ln{x +1 \over x-1}~~~,
\eeq
and $m_D^2 = N_f {g^2 \mu^2 \over 2 \pi^2}$ is the Debye screening mass.
The small x expansion of G leads to the Landau damped magnetic
gluon propagator
\beq
\label{LDD}
D^T_{\mu \nu} (q_0, q) 
~=~ { P^T_{\mu \nu} \over q^2 + i\frac{\pi}{2} m_D^2 \frac{|q_0|}{q} }~~~,
\eeq
while the expansion of F leads to the usual longitudinal propagator, with
Debye screening. 
The effect of Landau damping is to cut off the small-q divergence in
(\ref{LDD}) at $q \sim q_0^{1/3} m_D^{2/3}$.

One must be careful to compute the screening
effects using a Wilsonian cutoff. When we integrate over a shell
in momentum space only the contribution of quarks which have already 
been integrated out is to be included in the screening effects.
This means that we must re-examine the calculation which leads
to the Debye mass and Landau damping, and use cutoff-dependent
versions of the vacuum polarization 
$\equiv \Pi_{\mu \nu}^{\Lambda} $  in our RG. 

The terms in F and G can be shown to result from integration of
quark modes within roughly $q$ of the FS. For example, the Landau damping term
originates from an integral of the form 
(arising from (\ref{Pi}) \cite{Bellac}): 
\beq
\label{Pint}
I(\omega, \vec{q}) ~=~ \int dk~ d\Omega ~\frac{k^4}{E_1 E_2} ~
\Biggl[~ 
  \frac{  n ( E_1 ) - n ( E_2 )}{i \omega - E_1 + E_2} ~\Biggr]~~~,
\eeq
where $E_1 = k$ , $E_2 = | \vec{k} - \vec{q} | \simeq k - q \cos \theta$ and 
\beq
n (E) = {1 \over e^{ \beta (E - \mu)} + 1 }
\eeq
is the Fermi-Dirac distribution for quarks. $n (E)$ is
a theta function of $(\mu - E)$ in the low temperature limit
in which we work. For a given value of 
$\theta = \frac{\vec{k} \cdot \vec{q}}{k q}$,
the modes which contribute to Landau damping must be within 
$\simeq q \cos \theta$ of the FS due to the theta functions.

If the momentum transfer $q < \Lambda << \mu$, then none of these
modes are included in the calculation of $\Pi_{\mu \nu}^{\Lambda}$.
The corresponding 
$F^{\Lambda}$ and $G^{\Lambda}$ 
are changed drastically: in particular the leading 
$\mu^2$ behavior of $F$ vanishes,
and $G$ becomes proportional to $\mu^2$. 
Fortunately, our interest is only in the logarithmic divergences
of diagrams, which are dominated by 
gluon momenta satisfying the limit $\Lambda << q << \mu$, due to the
form of Landau damping. (i.e., $\Lambda \sim \Delta$, while the
dominant momentum transfer is $q \sim \Delta^{1/3} m_D^{2/3}$.)
In this limit the results for F an G given
in (\ref{FG}) are accurate.

Finally, we mention the issue of screening due to the diquark 
condensate itself, which is necessary for a self-consistent 
description of the region near the FS \cite{HongSD,SWgap}. In
a conventional 
superconductor the relative size of the magnetic penetration
depth $\lambda$ and the correlation length $\xi \sim \Delta^{-1}$
 determine whether one is in the type I ($\xi >> \lambda$) 
or type II ($\lambda << \xi$) regime. In a type II superconductor
the magnetic screening length can be computed using the London 
formula, and is proportional to the total density of superconducting
particles. In our case this would lead to a rather large effective 
screening mass $\lambda^{-1} \sim g^2 \mu^2$
relative to the gap size. However, in high density quark matter we
are actually in the type I, or Pippard, regime. In this regime
the effective screening mass is much smaller, and scales with the
gap $\Delta$.
A direct calculation of the gluon vacuum polarization using the
quark propagator in the presence of a gap  
(see (\ref{gapprop}) below) shows that the London-type
screening applies only to long wavelength gluons with momentum less 
than $\sim \Delta$. Harder gluons experience a much smaller screening
of the Pippard type. It is easy to see that these effects are too
small to affect our RG calculation; their contributions are dominated 
by Landau damping effects.

\section{RG Equations}

\begin{figure}[htb]
\center{
\leavevmode
\epsfbox{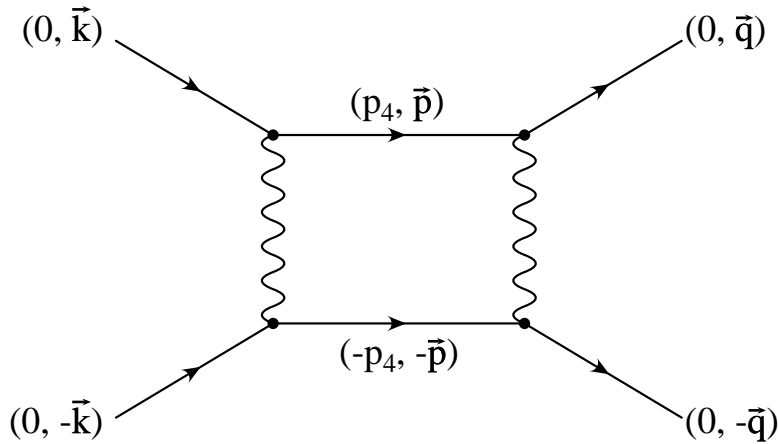}
\caption{Box diagram.} \label{boxdiagram}
}
\end{figure}

To obtain the RG equations we need to evaluate the scale-dependent
quantum corrections to the form-factor $G(k-q)$.
Let us consider the effect of integrating out quark states in 
the momentum shell $\Lambda' < |\vec{q}| - \mu < \Lambda$, and
gluon states in the momentum shell $\Lambda' < |\vec{q}| < \Lambda$. 
The tree level contribution comes directly 
from one-gluon exchange, figure \ref{gluon}, while
the one loop contribution comes from the box diagram, 
figure \ref{boxdiagram}. 
It is important to note that only one topology
of the box diagram contributes. The diagram with ``crossed'' gluon lines
does not have the same kinematic structure as an iterated Cooper pairing
interaction, and is subleading. Actually, in our effective theory
the box diagram contains several different contributions. 
The most important contains two local four fermion interactions, 
and is actually a ``bubble''
diagram, with running form factor coefficients. These coefficients contain
the effects of previously integrated shells of quarks and gluons. 
The other contributions involve
at least one exchange of a soft gluon within the momentum shell, and are
suppressed in the thin-shell limit. Thus, as we discuss below, the result
of the one loop part of our calculation is essentially the same as 
iterating bubbles with form factor vertices. 

Solving the RG equations is
equivalent to summing up an infinite series of ladder graphs corresponding
to nearly colinear scattering mediated by gluon exchange. This corresponds
to the ``rainbow'' approximation in which the gap equation is solved in
section 5.

\epsfysize=6 cm

Let us elaborate on why the tree and box graphs are sufficient to compute
the leading order solution to the RG equations. Consider integrating
out all modes between an upper cutoff $\sim \mu$ and a lower
cutoff given by $\Delta$. For simplicity, we can consider doing this
in a single step, rather than shell by shell. A systematic
expansion for the result is possible in powers of the gauge coupling
g and powers of $t \sim \ln \Lambda $. Since the RG evolution terminates
at a Landau pole of order $\sim \exp (-c/g)$, $t \sim 1/g$ in our
counting scheme. As we shall see below,
(figure \ref{gluon}) 
is of order $g^2 t$, and so contributes a term of order $g^2$ to the
RG equation. The box diagram (figure \ref{boxdiagram}) 
is of order $g^4 t^3$ and so contributes
terms of order $g^4 t^2 \sim g^2$. Higher order loop corrections to
either of these graphs (e.g. from vertex or wavefunction renormalization)
are suppressed by at least $g^2 t \sim g$. Of course, as discussed in
the previous section, the screened gluon propagators must be used in
our computations, since some of the logarithms which arise are due
to small angle scattering. Additional radiative corrections 
beyond vacuum polarization effects on the gluon propagators 
are suppressed by at least a power of $g^2 t$.

Because the rotational SO(4) symmetry is broken to SO(3), the
coupling in (\ref{G}) will split into its ``temporal'' $G_4$ and 
``spatial'' $G_i$ parts, which we refer to as A and B components,
according to the notation previously used in \cite{EHS}
\beqa
A (k-q) \bar{\psi_+} (k_0, k) \gamma_{4} P_L \psi_+ (q_0,q) 
 \bar{\psi}_+ (k_0, -k) \gamma^{4} P_L \psi_+ (q_0, -q) ~+~ \nonumber \\
B (k-q) \bar{\psi}_+ (k_0, k) \gamma_{i} P_L \psi_+ (q_0,q) 
 \bar{\psi}_+ (k_0, -k) \gamma^{i} P_L \psi_+ (q_0, -q)~~~.~~~
\eeqa

Due to the form of the screening  effects described in 
the previous section it is easy to see that the RG equations for the 
$A$-type couplings will not contain terms of order $g^2$. 
The Debye mass $m_D$ in the longitudinal 
part of the gluon propagator (\ref{D}) removes any IR logarithm from
the one gluon exchange diagram.

In contrast, the $B$ coupling may recieve such corrections. The
leading contribution will come from the region of shell integration 
where one may neglect the Landau damping term $G(Q)$ in gluon propagator
(\ref{D}). In this region the scattering angle satisfies the following 
condition: $1 - x^{2/3} ~<~ z \, = \, cos \, \theta ~<~ 1 - x'^{2/3}$, with
$x = {\pi \over 2^{5/2}} {m_D^2 \Lambda \over \mu^3}$. The angular
momentum $l$-component $B_l$ receives the contribution
\beq
\label{Btree} 
B_l (\Lambda) ~=~ -\,{1 \over 3}{2 \over 3} {g^2 \over 4 \mu^2} 
  \int_{-1}^{1-x^{2/3}} {dz \over 1 - z} P_l(z)
\eeq
(the factor of $1/3$ 
corresponds to the attractive $\bar{3}$ channel, while the factor of
$2/3$ arises from the transversality of the magnetic gluon propagator).
The correction is of the form
$\delta B_l \sim {g^2 \over \mu^2}\, \delta t$,
proportional to the RG scaling parameter $\delta t = 
\ln {\Lambda \over \Lambda'}$.
Note that  the leading term in $\delta B_l$ 
does not depend on the details of the angular cut-offs
-- that ambiguity is cancelled in the definition of $t$.

The box diagram of figure \ref{boxdiagram} has the following integral 
representation\footnote{For notational simplicity, 
we suppress the chiral projectors
$P_{L}$ in our expressions, although they appear in our
calculations.}:
\beq
\label{box}
I(t) ~=~ - g^4 \int {d^4 p \over (2 \pi)^4}\, 
K_{\mu\rho, \nu\kappa}\,
\gamma^\mu 
\Big[(i \mu +p_4) \gamma^4 + \vec{p} \cdot \vec{\gamma}\Big] \gamma^\nu
\otimes \gamma^\rho
\Big[(i \mu - p_4) \gamma^4 - \vec{p} \cdot \vec{\gamma}\Big] \gamma^\kappa ~,
\eeq
where $K_{\mu\rho, \nu\kappa}$ incorporates both fermion and
propagators (\ref{D}):
\beq
\label{K}
K_{\mu\rho, \nu\kappa} ~=~ 
{D_{\mu\rho} (p_4,\, \vec{k} - \vec{p}) 
\over (i \mu + p_4)^2 + \vec{p}\,^2 } \, 
{D_{\nu\kappa} (p_4,\, \vec{p} - \vec{q}) 
\over (i \mu -p_4)^2 + \vec{p}\,^2 }~.
\eeq
In order to evaluate the contribution to I(t) one should apply the familiar 
decomposition of momenta near FS: $\vec{p} = \vec{\mu}_p + \vec{p}_n,
\vec{k} = \vec{\mu}_k, \vec{q} = \vec{\mu}_q$ with normal component being 
integrated in the limits $\Lambda'^2 < p_n^2 < \Lambda^2$ \cite{EHS}.
Note that due to screening effects the dominant 
regions of $p_n$ and $p_4$ are around the origin, close to the FS.

Consider, for example, the part of $I(t)$ with both gluon lines being 
transverse. All other cases are done analogously. Capturing only the leading 
$O(\ln^3\Lambda)$ behavior one gets
\beq
\label{Tbox}
I(t)_T ~\simeq~ {g^4 \over 16 \mu^2} \int \int {d p_4 d p_n \over (2\pi)^4}
\int d\Omega_p \,
{ \gamma_i \gamma^4 \gamma_j \otimes \gamma^i \gamma^4 \gamma^j  ~+~
\gamma_i (\vec{p}/\mu \cdot \vec{\gamma}) \gamma_j \otimes
 \gamma^i (\vec{p}/\mu \cdot \vec{\gamma}) \gamma^j
\over
(p_4^2 ~+~ p_n^2) (1~-~ z_1) (1~-~ z_2)}~,
\eeq
where $z_1$ is the cosine of the
angle between $\vec{\mu_k}$ and $\vec{\mu_p}$, and
$z_2$ is the cosine of the angle between $\vec{\mu_q}$ and $\vec{\mu_p}$.
Then, use the definition (\ref{Btree}) of $B_l$ in the
expansion of transverse gluon propagators 
\beq
{1 \over 1~-~ z} ~=~ - {12 \mu^2 \over g^2} \sum_l\, (2 l + 1)\, B_l 
\, P_l(z)
\eeq
on the interval\footnote{
The longitudinal gluon propagators may be expanded in Legendre 
polynomials $P_l(z)$ on the interval $-1< z < 1$.}
$1 < z < 1 - x^{2/3}$.
Applying orthogonality conditions\footnote{The averaging over coordinate
systems is assumed in the second orthogonality condition of
(\ref{ortho}).} :
\beqa
\label{ortho}
\int d \Omega ~P_l\, (z_1)\, P_{l'} (z_2)~&=&~
\delta_{ll'} \, {4 \pi \over 2l +1} \, P_l(z)~ \nonumber \\
\int d \Omega ~\hat{p_i} \hat{p_j} ~P_l\, (z_1)\, P_{l'} (z_2)~&=&~
\delta_{ll'} \, {\delta_{ij}\over 3} \, {4 \pi \over 2l +1} \, P_l(z)~,
\eeqa
one obtains the final answer for the $I(t)_T$ in the leading divergence 
approximation
\beq
\label{T1box}
I(t)_T ~\simeq~ - {\mu^2\over 4 \pi^2} \Big( 5\,
\gamma_4 \otimes \gamma^4 ~+~ {13 \over 3}\, \gamma_i \otimes \gamma^i \Big)
\, B_l^2(\Lambda)\, t~.
\eeq
The final answer for the (Minkowskian) RGE equations is
\beqa
\label{AB}
{d A_l \over dt} ~&=&~ - \, {N \over 2} 
\Big( A_l^2 ~-~ 2 A_l B_l ~+~ 5 B_l^2 \Big)\, ~~~~~~~~~~~~~~~ \nonumber \\
{d B_l \over dt} ~&=&~  {N \over 6} 
\Big( A_l^2 ~-~ 10 A_l B_l ~+~ 13 B_l^2 \Big) ~+~ {g^2 \over 27 \mu^2}~, 
\eeqa
where $N= \mu^2/2 \pi^2$.
These  RG equations are the same as obtained previously for the case of local 
interactions \cite{EHS}, except for the constant 
term in the $B_l$ sector. A simple way to understand this is as 
follows:
by expanding the four quark form factor in terms of orthogonal local operators
$G_l$, we reduce the problem to one in which the gluon exchanges are
replaced by local interactions. Hence the analysis of \cite{EHS}
should apply and the same RG equations obtained for each value of l.
This is true as well for LR helicity scattering, which we do not
consider explicitly here. 
Note that though the $A_l$ couplings originate in the Debye screened sector, 
it would be a mistake to discard them in the RGE equations 
due to the fact that there is mixing with the $B_l$, which diverge near
the FS.
Diagonalization of (\ref{AB}) gives
\beqa
\label{ABdiag}
\frac{d}{dt} \Big(A_l ~-~ 3 B_l\Big)  ~&=&~ - \, N 
\Big(A_l ~-~ 3 B_l\Big)^2 ~-~ {g^2 \over 9 \mu^2}\, \nonumber \\
\frac{d}{dt} \Big(A_l~ +~ B_l\Big)  ~&=&~ - \,  {N \over 3}
\Big(A_l ~+~ B_l\Big)^2 ~+~ {g^2 \over 27 \mu^2}~.
\eeqa
It is straightforward to solve these equations. 
For convenience, we define the spin 0 combination
$S_l \equiv A_l - 3 B_l$ and the spin 1 combination
$T_l \equiv A_l + B_l$. We find that $S_l$ reaches a Landau pole 
at the scale
\beq
\label{LP1}
\Delta \sim \mu g^{-5} \exp \Bigl( - {3 \pi^2 \over \sqrt{2} g} \Bigr) ~~~,
\eeq
which is agrees with the result (\ref{LP}), despite differences between
the RG equations of \cite{Son} and ours (see below). 
$T_l$, due to the opposite sign of the $g^2$ term, does not diverge and
reaches the asymptotic value of 
$T_l (t \rightarrow \infty) = {\sqrt{2} \pi g \over 3 \mu^2}$.

In order to compare our results with those of \cite{Son}, it is
necessary to convert four quark operators into scattering amplitudes.
There are additional angular dependences introduced by the spin
angular momentum of the quarks. Let us classify amplitudes 
by their total angular momentum, which is the sum of the
spin and orbital components. Thus, for example, the j=0 channel 
receives contributions from both the l=0,s=0 and  l=1,s=1 
operators.  First we note that the spinor
part of operators of type A introduce an additional factor of 
$~2 \cos^2 \frac{\theta}{2}~$ to the scattering amplitude, while the 
type B operators introduce a factor of 
$~2 \cos^2  \frac{\theta}{2} + 4 \sin^2 \frac{\theta}{2}~$. This leads
to the following expression for the amplitude
\beq
\label{amp1}
f(\theta) = \sum_l (2l+1) \Bigl[ S_l + T_l \cos \theta
\Bigr] P_l ( \cos \theta )
\eeq
The components
of total angular momentum j are given by
\beq
\label{fj}
f_j = \Bigl[ S_j ~+~ \frac{j}{2j+1} T_{j-1} ~+~ \frac{j+1}{2j+1} T_{j+1}
\Bigr]~~~,
\eeq
where we have used the identity
\beq
(2l+1) z P_l (z) = (l+1) P_{l+1} (z) + l P_{l-1} (z)~~~.
\eeq
The RG equations for $f_j$ can be deduced easily from (\ref{fj})
and (\ref{ABdiag}). For the lowest component, we have
\beq
\label{f0}
\frac{d}{dt} f_0 ~=~ - \frac{2 g^2}{27 \mu^2} - N S_0^2 - \frac{N}{3} T_1^2~~~.
\eeq
The equations for higher components are of the form
\beq
\label{fl}
\frac{d}{dt} f_j ~=~ - \frac{2 g^2}{27 \mu^2} -N S_j^2 - \frac{N}{3}
\Bigl( \frac{j}{2j+1} T_{j-1}^2 + \frac{j+1}{2j+1} T_{j+1}^2 \Bigr)~~.
\eeq
Several remarks are in order.

\noindent $\bullet$ 
These results are similar to, but different from, those of
\cite{Son}. In particular, we do not find that the RG equations take 
the simple form \cite{Son}
\beq
\frac{d}{dt} f_j ~=~ - \frac{g^2}{9 \mu^2} - N f_j^2~~~.
\eeq
The rhs of our equation (\ref{fl}) cannot be organized in terms
of any simple amplitude $f_j^2$.
We believe that the treatment of quantum corrections in
section 2 of that paper is too crude. In particular, iterating the
amplitude $f(p,k)$ in order to obtain the $f_j^2$ terms in the RG
equation neglects some important spinor structure of the vertex
which our calculation takes into account.

\noindent $\bullet$ Fermi statistics constrain the operators as
follows.  We restrict ourselves to the color $\bar{3}$ and isospin singlet
channels, which are both antisymmetric. The remaining part of the 
wavefunction must be
antisymmetric. In the antisymmetric $s=0$ channel, we must therefore
have $l = 0,2,4,...$, while in the symmetric $s=1$ channel we must
have $l=1,3,5,...$. Thus the operators $S_l$ vanish for odd l and $T_l$
vanish for even l. From (\ref{fj}) we see that $f_j$ vanishes for all
odd values of j. Note that this analysis is modified in the LR channels,
which we do not consider here.

\noindent $\bullet$
It is not necessary to solve the $f_j$ RG equations, as their
behavior can be deduced from that of $S_l$ and $T_l$. Since $T_l$
never diverges, near the FS $f_j \approx S_j$ and diverges at the
scale (\ref{LP1}). Note that the $\frac{N}{3} T_j^2$ term cannot 
be neglected in (\ref{fl}); 
its asymptotic value is ${g^2 \over 27 \mu^2}$.

\noindent $\bullet$ 
Because the Landau poles in all the j channels are the same,
we naively expect to find condensates with non-zero angular momentum,
leading to the breaking of rotational symmetry \cite{Son}.
However, we will see in the next section by studying the gap
equation that this is not the case.

\section{Rotational Symmetry Breaking and Gap Equation}

In this section we explore the issue of condensates with
non-zero angular momentum
using the gap equation. Our RG results suggest that the
gap function could violate rotational invariance.
Some recent papers
have studied the s-wave condensate using the non-local gap
equation, including the effects of magnetic gluons 
\cite{HongSD,SWgap,PRTc}. We will generalize their approach to consider
gap functions $\Delta (q_0, \vec{q})$ which are functions of
orientation. Remarkably, it is possible to show that in the leading
order approximation only an s-wave condensate is formed.

Let us introduce a two component field 
$\Psi=(\psi,\bar\psi^T)$.
The inverse quark propagator takes the form
\beq
\label{gapprop}
S^{-1}(q) = \left(\begin{array}{cc}
q\!\!\!/  +  \mu\!\!\!/ - m &  \bar\Delta \\
 \Delta  & (q\!\!\!/  +  \mu\!\!\!/ + m)^T 
\end{array}\right),
\eeq
where $\bar\Delta=\gamma_0\Delta^\dagger\gamma_0$. 
The gap is a matrix in color, isospin, and Dirac space, and in
our analysis we allow it to depend on orientation. As discussed,
the RG analysis shows that the condensate will form in the LL (RR)
channels. Given this, the form of the gap matrix is
\cite{colorsuper,PR}
\beq
 \Delta^{ab}_{ij}(q) = (\lambda_2)^{ab}(\tau_2)_{ij}
  C\gamma_5 \left( \Delta_+ (q_0,\vec{q} )\frac{1}{2}(1+\vec\alpha\cdot\hat q )
    +\Delta_- (q_0, \vec{q} )\frac{1}{2}(1-\vec\alpha\cdot\hat q ) \right).
\eeq
In our present weak coupling discussion, we are only interested in $\Delta_+$. 
$\Delta_-$ corresponds to a condensate of anti-quarks, and does not
influence $\Delta_+$.
Henceforth we shall only refer to $\Delta \equiv \Delta_+$.
Making the usual FS approximations, the gap equation has the
form
\beq
\Delta (q_0, q, \hat{q} ) =
  - i g^2 \int {d^4 k \over (2 \pi)^4} ~ { D (q_0 - k_0 , \vec q - \vec k) ~
\Delta(k_0, k, \hat{k}) \over \epsilon(k)^2 + k_0^2 + 
|\Delta(k_0 ,k, \hat{k})|^2     }
\label{gapV}
\eeq
where the interaction kernel $D(q_0 - k_0, \vec q - \vec k)$ is 
essentially the gluon propagator up to some additional factors arising
from the particle projector and gamma matrices. 
Here $k = | \vec{k}|$, $q = |\vec{q}|$
and $\epsilon (k) = k - \mu~$; we assume $q_0^2 \sim q_0^2 << \mu^2$
and $k \sim q \simeq \mu$. 
For the next step in our analysis it
is useful to separate the interaction kernel into angular momentum
channels,
\beq
D(q_0 - k_0, \vec q - \vec k) = \sum_{l} (2 l + 1) 
P_l(\cos \theta_{kq} ) ~ D^l(q_0 - k_0, | \vec{q} - \vec{k} |)~~~.
\eeq
The $D^l$ coefficients are obtained by integrating the kernel against
Legendre polynomials $P_l ( \cos \theta )$. Due to the divergence at
small angle, all of the $D^l$ have the same value at leading logarithmic
order. 

The gap equation can now be rewritten as
\beq
\Delta (q_0, q, \hat{q} )
 = - 4 \pi i g^2 \sum_{l m}  Y^{l}_m(\theta_q, \phi_q)  ~
\int {d ^4 k \over (2 \pi)^4}~ Y^{l *}_m(\theta_{k}, \phi_{k}) ~
~ {D^{l}(q_0 - k_0, | \vec{q} - \vec{k} | )
 ~ \Delta( k_0,k,\hat{k} )
\over \epsilon ^2 (k) + k_0^2  + |\Delta (k_0,k,\hat{k})|^2 } ~~ .
\eeq
For $D^l = D (q_0 - k_0, | \vec{q} - \vec{k} | ) $ independent of l, 
the sum over spherical harmonics reduces to a delta function: 
$\delta ( \phi_q - \phi_k ) ~\delta ( \cos \theta_q ~-~ \cos \theta_k )$.
The angular integral on the rhs of the equation can be trivially evaluated,
yielding
\beq
\label{gap1}
\Delta (q_0, q ,\hat{q})  
= -i g^2 \int \frac{dk_4 dk k^2}{(2 \pi)^3 }~ 
{ D (q_0 - k_0, | \vec{q} - \vec{k} | )    \Delta(k_0, k,\hat{q}) 
\over 
\epsilon ^2 (k) + k_0^2  + |\Delta(k_0,k,\hat{q})|^2   }  ~~~.
\eeq
This equation can be regarded as a set of identical equations, one for
each orientation $\hat{q}$. Since each equation is identical, the solution
must be independent of orientation.

This result could have been guessed directly from the fact that colinear
scattering dominates the magnetic effects. In that approximation the kernel
$D$ is proportional to a delta function $\delta (\hat{k} - \hat{q})$, and
it is clear that (\ref{gapV}) only has an s-wave solution. 

Our analysis thus far has been within the leading logarithmic approximation.
We can relax this condition by considering a gap function
\beq
\label{exp}
\Delta (q) = \Delta^0 + \Delta^1 (q_0,\vec{q})~~~,
\eeq
where $\Delta^0$ is the constant (in orientation) 
solution obtained from (\ref{gapV}),
and $\Delta^1$ is a small perturbation which can depend on orientation.
We will show that $\Delta^1$ is at most of order $\exp( - c / g^2)$ and hence
negligible relative to the $\Delta^0$. Substituting (\ref{exp}) into
(\ref{gapV}), we obtain two gap equations. The leading order equation
is the just the usual one for an s-wave condensate, and determines 
$\Delta^0$. Note that we retain the complete interaction kernel
$D(q_0 - k_0, \vec q - \vec k )$ here, without making the
leading log approximation.
The second equation contains terms of ${\cal O} ( \Delta^1 )$:
\beq
\label{gapD1}
\Delta^1 (q_0, \vec{q}) ~\simeq~   g^2 \mu^2 \int 
\frac{dk_0 d\Omega_{kq}}{(2 \pi)^3} 
{D ( q_0 - k_0, \theta_{kq}  ) \over 2 \sqrt{ k_0^2 + | \Delta^0 |^2}}~
\Biggl[
1 - \frac{| \Delta^0|^2}{ k_0^2 + | \Delta^0 |^2}
\Biggr]~ \Delta^1 (k_0, \vec{k})~~~.
\eeq
This equation was obtained after first performing the integral over k,
in the approximation that $\Delta$ and $D$ are slowly varying for 
$k \simeq \mu$.
The term in brackets in (\ref{gapD1}) suppresses
the logarithmic divergence in the integral over $k_0$ near the FS,
although there is still a potential divergence from the small angle
behavior of $D ( q_0 - k_0, \theta_{kq}  )$. Hence
the solution $\Delta^1$ is at most of order $\exp(-1/g^2)$,
and is negligible relative to $\Delta^0$ in the weak coupling limit. 

The discussion in terms of $\Delta^0$ and $\Delta^1$ is
quite general. We can also apply it to lower densities,
where the magnetic interaction
is presumably screened by non-perturbative effects, 
and quark interactions are effectively local. In this case the 
largest condensate can be shown to
be rotationally invariant \cite{colorsuper,EHS}. Because
$D$ in (\ref{gapD1}) is non-singular, a non-zero solution $\Delta^1$
only exists above some critical coupling (if at all). This is unlike the usual 
Cooper pairing instability in which an arbitrarily weak interaction can 
lead to a condensate. It seems that a large s-wave
component tends to inhibit condensates of higher angular momentum.

Having made some general observations about rotational invariance,
we now concentrate on solving the gap equation explicitly, in order to
check our result for the Landau pole (\ref{LP1}). The authors of 
\cite{HongSD,SWgap,PRTc} do not consider higher orbital angular momentum
components $\Delta_{l > 0}$ in their analyses, but it is straightforward
to do so.

The explicit gap equation for $\Delta_1 \equiv \Delta$, neglecting the
suppressed antiparticle contribution, is \cite{SWgap,PRTc}
\beq
\label{detailgap}
\Delta (q_0, \vec{q}) ~=~ - \frac{2ig^2}{3} \int {d^4k \over (2 \pi)^4}
{\Delta( k_0, \vec{k} ) \over 
\epsilon ^2 (k) + k_0^2  + |\Delta (k_0,\vec{k}) |^2 }
{ \frac{3}{2} - \frac{1}{2} \hat{k} \cdot \hat{q} \over
(k - q)^2 + G}~~~.
\eeq
Here we have written explicitly the factors resulting from the 
particle projector and gamma matrices. G is the Landau damping term
which appears in the magnetic gluon propagator; we neglect
the effect of the electric gluon.

In order to consider higher angular modes of $\Delta$ we make the
expansion:
\beq
\Delta (q_0, \vec{q}) = \sum_l (2l+1) P_l (\hat{q}) \Delta_l~~~,
\eeq
as well as a similar expansion of the gluon propagator with
coefficients
\beq
D_l = \frac{1}{2 \mu^2} \int_{-1}^{1} dz 
{ P_l (z) \over 1 - z + G/(2 \mu^2)}~~~,
\eeq
which are all of the same size at leading order. Using the identity
\beq
P_l ( \cos \theta_{kq} ) ~=~ \frac{2 \pi}{2l+1}
\sum_{m = -l}^{m = l}~ Y_l^m ( \theta_k, \phi_k )^* 
Y_l^m (\theta_q, \phi_q)~~,
\eeq
we obtain the following coupled gap equations
\beqa
\Delta_l &=& -    \frac{2ig^2}{3 \pi}  \int  {dk_0 dk \over (2 \pi)^2}~ 
 { \mu^2 \over \epsilon ^2 (k) + k_0^2  + |\Delta (k_0,\vec{k}) |^2 }~
\nonumber \\
&\times~&
\Bigl[ \frac{3}{2} D_l \Delta_l ~-~ \frac{1}{2(2l+1)} \Bigl(
(l+1) D_{l+1} \Delta_{l} ~+~ l D_{l-1} \Delta_{l} \Bigr) \Bigr]~~~.
\eeqa
In obtaining this equation we have neglected the angular dependence
of $|\Delta (k_0,\vec{k})|^2$ which occurs in the denominator. This
is justified if the gap turns out to be rotationally invariant, as
expected from our previous arguments. It is easy to see that a 
self-consistent solution exists with all $l > 0$ gaps zero, and the 
solution for $\Delta = \Delta_0$ given by (\ref{LP1}).

\section{Discussion}

In this paper we investigated the renormalization group behavior of
QCD at high density, concentrating on the effects of long range
magnetic interactions. Our approach was somewhat different from that
of \cite{Son} in that we focused on individual local operators rather 
than scattering amplitudes. The resulting RG equations are different, 
although the location of the Landau pole is still given by
(\ref{LP}). The disagreement results from two causes: (I) the transverse
form of the propagator does not appear to have been used in \cite{Son},
leading to a different coefficient in the constant $g^2$ term of the RGE
and (II) the treatment of spinor properties of the quantum corrections
is different in the two calculations.
We believe that the renormalization group applied directly to amplitudes
does not properly compute the loop corrections.
Some additional issues we attempted to clarify include the 
validity of the use of 
Landau damping in resummed gluon propagators, the gauge invariance
of the computation and the size of subleading corrections.
Our RG equations (\ref{ABdiag}) are gauge invariant, and represent
the leading order result in a self-consistent expansion. Corrections
to the coefficients in (\ref{ABdiag}) are of order ${\cal O} (g)$ in
the weak coupling limit.

We also used the gap equation to investigate whether rotational symmetry is
broken at asymptotically high densities. The gap equation analysis shows 
that scattering which is predominantly colinear leads to a 
rotationally invariant condensate. We found, in disagreement with
a naive interpretation of the RG results, 
that any condensates of higher angular momentum are
exponentially smaller than the s-wave condensate. We also checked that
the solution of the gap equation agrees with the value of our Landau pole.


\bigskip
\noindent 
The authors would like to thank Nick Evans, Deog-Ki Hong,
James Hormuzdiar, Rob Pisarski and Dirk Rischke 
for useful discussions and comments. 
This work was supported in part under DOE contracts 
DE-FG02-91ER40676 and DE-FG06-85ER40224.


\end{document}